\documentclass [8pt]{article}
\usepackage[T1]{fontenc}
\usepackage[final]{epsfig}

\begin{document}


%
%
%
\edef\catcodeat{\the\catcode`\@ }     \catcode`\@=11
\newbox\p@ctbox                       
\newbox\t@mpbox                       
\newbox\@uxbox                        
\newbox\s@vebox                       
\newtoks\desct@ks \desct@ks={}        
\newtoks\@ppenddesc                   
\newtoks\sh@petoks                    
\newif\ifallfr@med  \allfr@medfalse   
\newif\if@ddedrows                    
\newif\iffirstp@ss  \firstp@ssfalse   
\newif\if@mbeeded                     
\newif\ifpr@cisebox                   
\newif\ifvt@p                         
\newif\ifvb@t                         
\newif\iff@nished    \f@nishedtrue    
\newif\iffr@med                       
\newif\ifj@stbox     \j@stboxfalse    
\newcount\helpc@unt                   
\newcount\p@ctpos                     
\newdimen\r@leth      \r@leth=0.4pt   
\newdimen\x@nit                       
\newdimen\y@nit                       
\newdimen\xsh@ft                      
\newdimen\ysh@ft                      
\newdimen\@uxdimen                    
\newdimen\t@mpdimen                   
\newdimen\t@mpdimeni                  
\newdimen\b@tweentandp                
\newdimen\b@ttomedge                  
\newdimen\@pperedge                   
\newdimen\@therside                   
\newdimen\d@scmargin                  
\newdimen\p@ctht                      
\newdimen\l@stdepth
\newdimen\in@tdimen
\newdimen\l@nelength
\newdimen\re@lpictwidth
\def\justframes{\global\j@stboxtrue}  
\def\picturemargins#1#2{\b@tweentandp=#1\@therside=#2\relax}
\def\allframed{\global\allfr@medtrue} 
\def\emptyplace#1#2{\pl@cedefs        
    \setbox\@uxbox=\vbox to#2{\n@llpar
        \hsize=#1\vfil \vrule height0pt width\hsize}
    \e@tmarks}
\def\boxplace{\pl@cedefs\afterassignment\re@dvbox\let\n@xt= }
\def\re@dvbox{\setbox\@uxbox=\vbox\bgroup
         \n@llpar\aftergroup\e@tmarks}
\def\fontcharplace#1#2{\pl@cedefs     
    \setbox\@uxbox=\hbox{#1\char#2\/}%
    \xsh@ft=-\wd\@uxbox               
    \setbox\@uxbox=\hbox{#1\char#2}%
    \advance\xsh@ft by \wd\@uxbox     
    \helpc@unt=#2
    \advance\helpc@unt by -63         
    \x@nit=\fontdimen\helpc@unt#1%
    \advance\helpc@unt by  20         
    \y@nit=\fontdimen\helpc@unt#1%
    \advance\helpc@unt by  20         
    \ifnum\helpc@unt<51
      \ysh@ft=-\fontdimen\helpc@unt#1%
    \fi
    \e@tmarks}
\def\n@llpar{\parskip0pt \parindent0pt
    \leftskip=0pt \rightskip=0pt
    \everypar={}}
\def\pl@cedefs{\xsh@ft=0pt\ysh@ft=0pt}
\def\e@tmarks#1{\setbox\@uxbox=\vbox{ 
      \n@llpar
      \hsize=\wd\@uxbox               
      \noindent\copy\@uxbox           
      \kern-\wd\@uxbox                
      #1\par}
    \st@redescription}
\def\t@stprevpict#1{\ifvoid#1\else    
   \errmessage{Previous picture is not finished yet.}\fi} 

\def\st@redescription#1\par{
    \global\setbox\s@vebox=\vbox{\box\@uxbox\unvbox\s@vebox}%
    \desct@ks=\expandafter{\the\desct@ks#1\@ndtoks}}
\def\def@ultdefs{\p@ctpos=1         
      \def\lines@bove{0}
      \@ddedrowsfalse               
      \@mbeededfalse                
      \pr@ciseboxfalse
      \vt@pfalse                    
      \vb@tfalse                    
      \@ppenddesc={}
      \ifallfr@med\fr@medtrue\else\fr@medfalse\fi
      }

\def\descriptionmargins#1{\global\d@scmargin=#1\relax}
\def\@dddimen#1#2{\t@mpdimen=#1\advance\t@mpdimen by#2#1=\t@mpdimen}
\def\placemark#1#2 #3 #4 #5 {\unskip    
      \setbox1=\hbox{\kern\d@scmargin#5\kern\d@scmargin}
      \@dddimen{\ht1}\d@scmargin        
      \@dddimen{\dp1}\d@scmargin        
      \ifx#1l\dimen3=0pt\else           
        \ifx#1c\dimen3=-0.5\wd1\else
          \ifx#1r\dimen3=-\wd1
     \fi\fi\fi
     \ifx#2u\dimen4=-\ht1\else          
       \ifx#2c\dimen4=-0.5\ht1\advance\dimen4 by 0.5\dp1\else
         \ifx#2b\dimen4=0pt\else
           \ifx#2l\dimen4=\dp1
     \fi\fi\fi\fi
     \advance\dimen3 by #3
     \advance\dimen4 by #4
     \advance\dimen4 by-\dp1
     \advance\dimen3 by \xsh@ft         
     \advance\dimen4 by \ysh@ft         
     \kern\dimen3\vbox to 0pt{\vss\copy1\kern\dimen4}
     \kern-\wd1                        
     \kern-\dimen3                     
     \ignorespaces}                    
\def\fontmark #1#2 #3 #4 #5 {\placemark #1#2 #3\x@nit{} #4\y@nit{} {#5} }
\def\fr@msavetopict{\global\setbox\s@vebox=\vbox{\unvbox\s@vebox
      \global\setbox\p@ctbox=\lastbox}%
    \expandafter\firstt@ks\the\desct@ks\st@ptoks}
\def\firstt@ks#1\@ndtoks#2\st@ptoks{%
    \global\desct@ks={#2}%
    \def\t@mpdef{#1}%
    \@ppenddesc=\expandafter\expandafter\expandafter
                        {\expandafter\t@mpdef\the\@ppenddesc}}
\def\testf@nished{{\let\s@tparshape=\relax
    \s@thangindent}}
\def\inspicture{\t@stprevpict\p@ctbox
    \def@ultdefs                  
    \fr@msavetopict
    \iff@nished\else\testf@nished\fi
    \iff@nished\else
      \immediate\write16{Previes picture is not finished yet}%
    \fi
    \futurelet\N@xt\t@stoptions}  
\def\t@stoptions{\let\n@xt\@neletter
  \ifx\N@xt l\p@ctpos=0\else                
   \ifx\N@xt c\p@ctpos=1\else               
    \ifx\N@xt r\p@ctpos=2\else              
     \ifx\N@xt(\let\n@xt\e@tline\else        
      \ifx\N@xt!\@mbeededtrue\else           
       \ifx\N@xt|\fr@medtrue\else            
        \ifx\N@xt^\vt@ptrue\vb@tfalse\else  
         \ifx\N@xt_\vb@ttrue\vt@pfalse\else 
          \ifx\N@xt\bgroup\let\n@xt\@ddgrouptodesc\else
           \let\n@xt\@dddescription 
  \fi\fi\fi\fi\fi\fi\fi\fi\fi\n@xt}
\def\e@tline(#1){\def\lines@bove{#1}
    \@ddedrowstrue
    \futurelet\N@xt\t@stoptions}
\def\@neletter#1{\futurelet\N@xt\t@stoptions} 
\def\@ddgrouptodesc#1{\@ppenddesc={#1}\futurelet\N@xt\t@stoptions}
\def\fr@medpict{\setbox\p@ctbox=
    \vbox{\n@llpar\hsize=\wd\p@ctbox
       \iffr@med\else\r@leth=0pt\fi
       \ifj@stbox\r@leth=0.4pt\fi
       \hrule height\r@leth \kern-\r@leth
       \vrule height\ht\p@ctbox depth\dp\p@ctbox width\r@leth \kern-\r@leth
       \ifj@stbox\hfill\else\copy\p@ctbox\fi
       \kern-\r@leth\vrule width\r@leth\par
       \kern-\r@leth \hrule height\r@leth}}
\def\@dddescription{\fr@medpict     
    \re@lpictwidth=\the\wd\p@ctbox
    \advance\re@lpictwidth by\@therside
    \advance\re@lpictwidth by\b@tweentandp
    \ifhmode\ifinner\pr@ciseboxtrue\fi\fi
    \createp@ctbox
    \let\N@xt\tr@toplacepicture
    \ifhmode                         
      \ifinner\let\N@xt\justc@py
      \else\let\N@xt\vjustc@py
      \fi
    \else
      \ifnum\p@ctpos=1               
        \let\N@xt\justc@py
      \fi
    \fi
    \if@mbeeded\let\N@xt\justc@py\fi 
    \firstp@sstrue
    \N@xt}
\def\createp@ctbox{\global\p@ctht=\ht\p@ctbox
    \advance\p@ctht by\dp\p@ctbox
    \advance\p@ctht by 6pt
    \setbox\p@ctbox=\vbox{
      \n@llpar                     
      \t@mpdimen=\@therside          
      \t@mpdimeni=\hsize             
      \advance\t@mpdimeni by -\@therside
      \advance\t@mpdimeni by -\wd\p@ctbox
      \ifpr@cisebox
        \hsize=\wd\p@ctbox
      \else
        \ifcase\p@ctpos
               \leftskip=\t@mpdimen    \rightskip=\t@mpdimeni
        \or    \advance\t@mpdimeni by \@therside
               \leftskip=0.5\t@mpdimeni \rightskip=\leftskip
        \or    \leftskip=\t@mpdimeni   \rightskip=\t@mpdimen
        \fi
      \fi
      \hrule height0pt             
      \kern6pt                     
      \penalty10000
      \noindent\copy\p@ctbox\par     
      \kern3pt                       
      \hrule height0pt
      \hbox{}%
      \penalty10000
      \interlinepenalty=10000
      \the\@ppenddesc\par            
      \penalty10000                  
      \kern3pt                       
      }%
      \ifvt@p
       \setbox\p@ctbox=\vtop{\unvbox\p@ctbox}%
      \else
        \ifvb@t\else
          \@uxdimen=\ht\p@ctbox
          \advance\@uxdimen by -\p@ctht
          {\vfuzz=\maxdimen
           \global\setbox\p@ctbox=\vbox to\p@ctht{\unvbox\p@ctbox}%
          }%
          \dp\p@ctbox=\@uxdimen
        \fi
      \fi
      }
\def\picname#1{\unskip\setbox\@uxbox=\hbox{\bf\ignorespaces#1\unskip\ }%
      \hangindent\wd\@uxbox\hangafter1\noindent\box\@uxbox\ignorespaces}
\def\justc@py{\ifinner\box\p@ctbox\else\kern\parskip\unvbox\p@ctbox\fi
  \global\setbox\p@ctbox=\box\voidb@x}
\def\vjustc@py{\vadjust{\kern0.5\baselineskip\unvbox\p@ctbox}%
      \global\setbox\p@ctbox=\box\voidb@x}
\def\tr@toplacepicture{
      \ifvmode\l@stdepth=\prevdepth  
      \else   \l@stdepth=0pt         
      \fi
      \vrule height.85em width0pt\par
      \r@memberdims                  
      \global\t@mpdimen=\pagetotal
      \t@stheightofpage              
      \ifdim\b@ttomedge<\pagegoal    
         \let\N@xt\f@gurehere        
         \global\everypar{}
      \else
         \let\N@xt\relax             
         \penalty10000
         \vskip-\baselineskip        
         \vskip-\parskip             
         \immediate\write16{Picture will be shifted down.}%
         \global\everypar{\sw@tchingpass}
      \fi
      \penalty10000
      \N@xt}
\def\sw@tchingpass{
    \iffirstp@ss                     
      \let\n@xt\relax
      \firstp@ssfalse                
    \else
      \let\n@xt\tr@toplacepicture
      \firstp@sstrue
    \fi  \n@xt}
\def\r@memberdims{\global\in@tdimen=0pt
    \ifnum\p@ctpos=0
        \global\in@tdimen=\re@lpictwidth
      \fi
      \global\l@nelength=\hsize
      \global\advance\l@nelength by-\re@lpictwidth
      }
\def\t@stheightofpage{%
     \global\@pperedge=\t@mpdimen
     \advance\t@mpdimen by-0.7\baselineskip 
     \advance\t@mpdimen by \lines@bove\baselineskip 
     \advance\t@mpdimen by \ht\p@ctbox      
     \advance\t@mpdimen by \dp\p@ctbox      
     \advance\t@mpdimen by-0.3\baselineskip 
     \global\b@ttomedge=\t@mpdimen          
     }
\def\f@gurehere{\global\f@nishedfalse
      \t@mpdimen=\lines@bove\baselineskip   
      \advance\t@mpdimen-0.7\baselineskip   
      \kern\t@mpdimen
      \advance\t@mpdimen by\ht\p@ctbox
      \advance\t@mpdimen by\dp\p@ctbox
      {\t@mpdimeni=\baselineskip
       \offinterlineskip
       \unvbox\p@ctbox
       \global\setbox\p@ctbox=\box\voidb@x
       \penalty10000   \kern-\t@mpdimen     
       \penalty10000   \vskip-\parskip      
       \kern-\t@mpdimeni                    
      }%
      \penalty10000                         
      \global\everypar{\s@thangindent}
      }
\def\s@thangindent{%
    \ifdim\pagetotal>\b@ttomedge\global\everypar{}%
      \global\f@nishedtrue             
      \else
        \advance\@pperedge by -1.2\baselineskip
        \ifdim\@pperedge>\pagetotal\global\everypar{}%
          \global\f@nishedtrue
        \else
          \s@tparshape                 
        \fi
        \advance\@pperedge by 1.2\baselineskip
      \fi}
\def\s@tparshape{\t@mpdimen=-\pagetotal
   \advance\t@mpdimen by\b@ttomedge    
   \divide\t@mpdimen by\baselineskip   
   \helpc@unt=\t@mpdimen               
   \advance \helpc@unt by 2            
   \sh@petoks=\expandafter{\the\helpc@unt\space}
   \t@mpdimeni=\lines@bove\baselineskip
   \t@mpdimen=\pagetotal
   \gdef\lines@bove{0}
   \loop \ifdim\t@mpdimeni>0.999\baselineskip 
     \advance\t@mpdimen  by \baselineskip
     \advance\t@mpdimeni by-\baselineskip
     \sh@petoks=\expandafter{\the\sh@petoks 0pt \the\hsize}%
   \repeat
   \loop \ifdim\b@ttomedge>\t@mpdimen         
     \advance\t@mpdimen by \baselineskip
     \sh@petoks=\expandafter{\the\sh@petoks \in@tdimen \l@nelength }%
   \repeat
   \sh@petoks=\expandafter
      {\the\sh@petoks 0pt \the\hsize}
   \expandafter\parshape\the\sh@petoks
   }

\descriptionmargins{2pt}
\picturemargins{15pt}{0pt}

\catcode`\@=\catcodeat        \let\catcodeat=\undefined

\emptyplace{3.3in \includegraphics{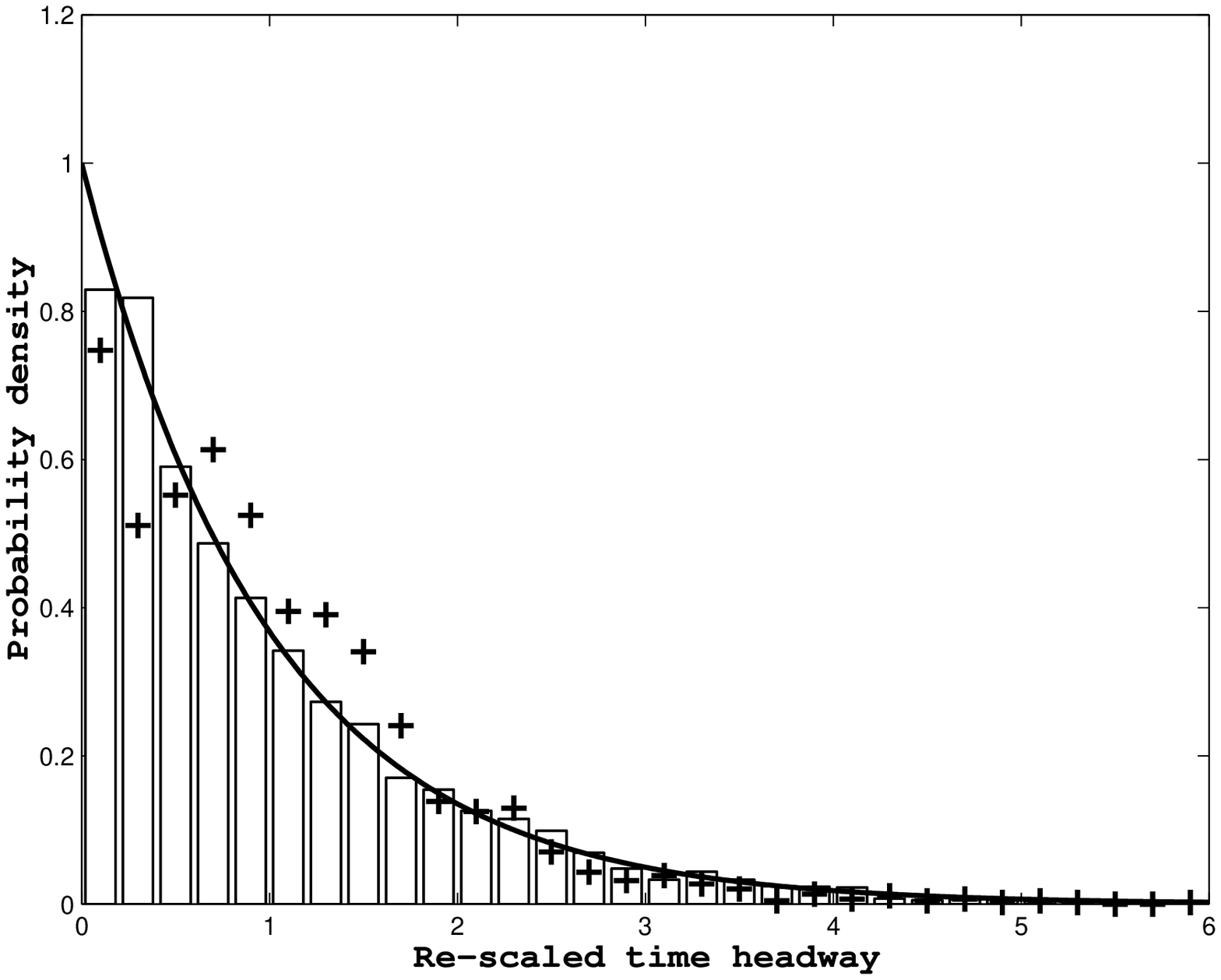}} {3.1in} {\footnotesize \noindent FIG.1. Time headway of the bus
transport in Puebla.\\ The curve represents the Poisson distribution
(\ref{Poisson}). Plus signs display the time headway distribution of buses in
Puebla and bars are taken from the CA model with $a=0.$ }

\emptyplace{3.3in \includegraphics{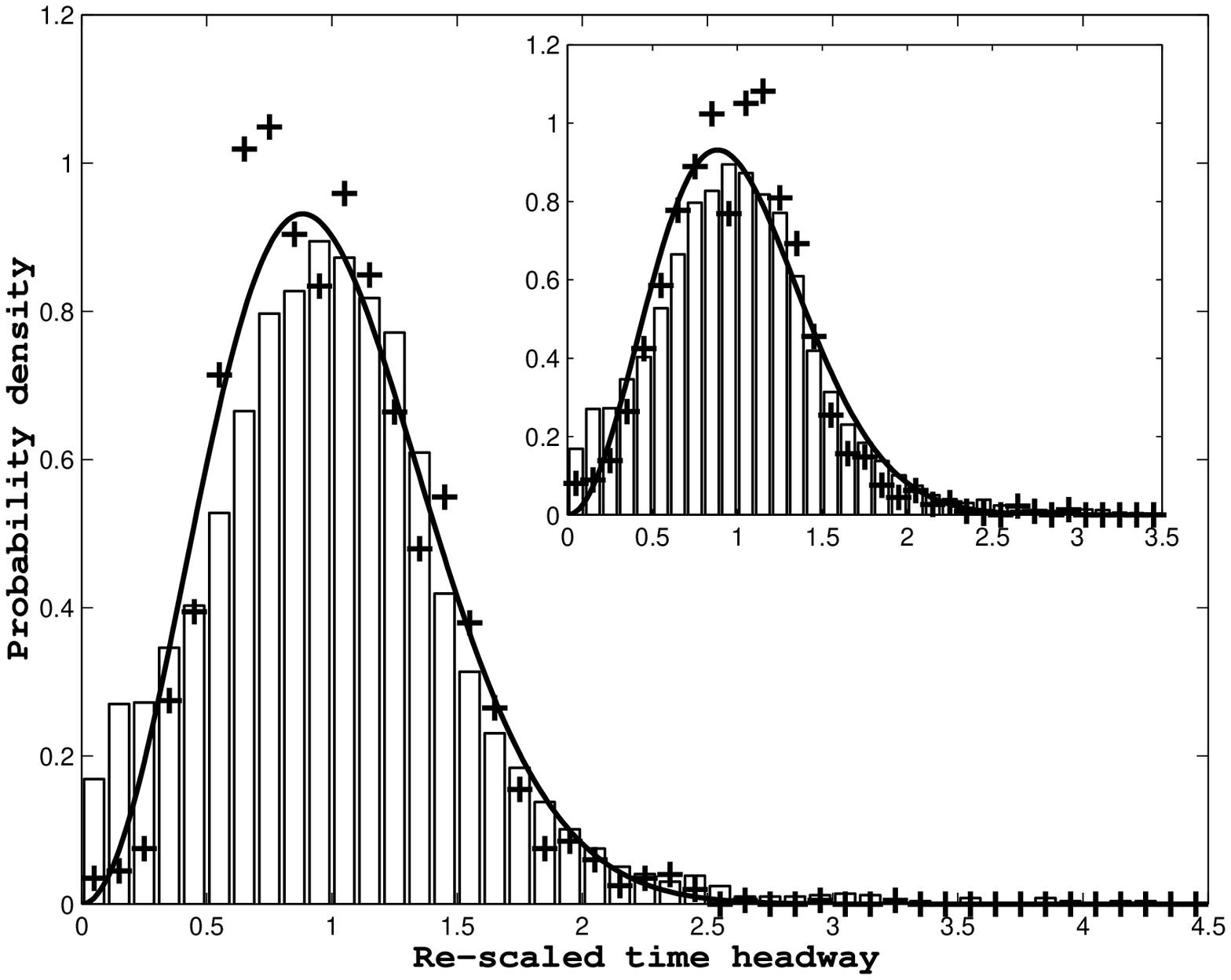}} {3.1in} {\footnotesize \noindent FIG.2. Time headway of
 bus transport in Cuernavaca and Mexico City.\\ The curve represents
the Wigner formula (\ref{Wigner}.) Plus signs display the time headway
distribution of buses in Cuernavaca and bars represent the results of the CA
model with $a=1/36.$ The results obtained for buses in Mexico City are
presented on the inset.}

\emptyplace{3.3in \includegraphics{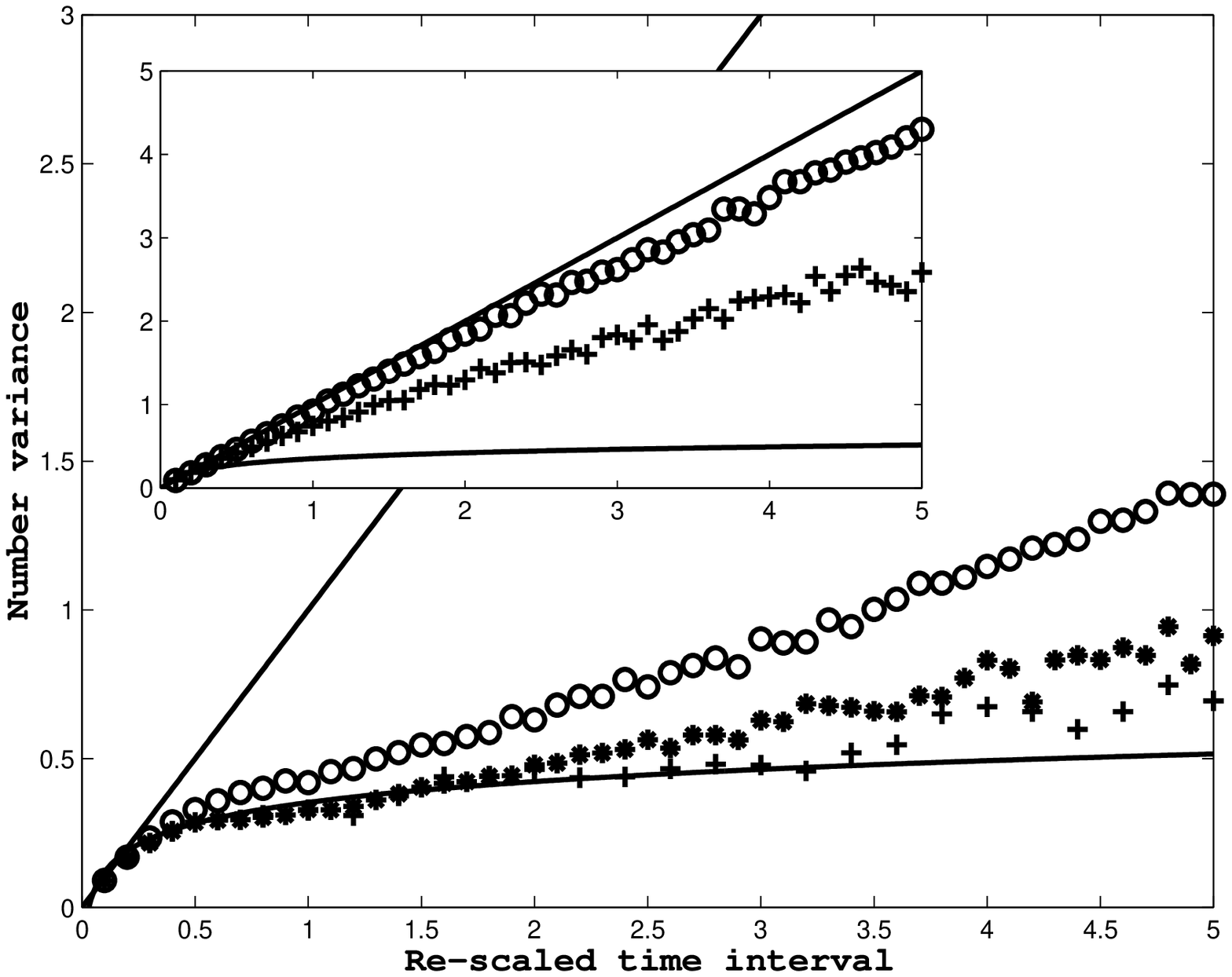}} {3.1in} {\footnotesize \noindent FIG.3. Number variance of the
public transport in some Mexican cities.\\ Curve and line represent the
prediction of the RMT (\ref{NV_GUE}) and (\ref{NV_Poisson}), respectively. Plus
signs and stars display number variance obtained from the public bus transport
in Cuernavaca and Mexico City respectively. Circles are the results of the CA
model for $a=1/36.$ The number variance obtained from the transport in  Puebla
is displayed on the insert (plus signs).The circles show the result of the CA
model obtained for $a=0.$}

\title{Headway statistics of public transport in Mexican cities}

\author{Milan Krbalek ${ }^{1,3}$ and  Petr Seba ${ }^{2,3}$\\
${ }^1$ \ Department of Mathematics, Faculty of Nuclear Sciences\\
and Physical Engineering,\\ Trojanova 13, Prague, Czech Republic\\
${ }^2$ \ Department of Physics, University  of Hradec
Kr\'alov\'e,\\ V\'{\i}ta Nejedl\'eho 573, Hradec Kr\'alov\'e,
Czech Republic\\ ${ }^3$ \ Institute of Physics, Czech Academy of
Sciences,\\ Cukrovarnick\'a 10, Prague, Czech Republic\\}

\maketitle

\begin{abstract}

We present a cellular automaton simulating the behavior of public
bus transport in several Mexican cities. The headway statistics
obtained from the model is compared to the measured time intervals
between subsequent bus arrivals to a given bus stop and to a
spacing distribution resulting from a random matrix theory.

\vspace*{3mm}
\end{abstract}


The public transport in Mexico is organized differently to that
known in Europe. First, no leading companies are responsible for
the city transport. Thus, there are no timetables for the city
buses and sometimes even not well defined bus stops. Moreover, the
driver is usually the owner of the bus and so his aim is to
maximize the income. Since every passenger entering the bus has to
pay, the driver tries to collect the largest possible number of
passengers. When not regulated the time interval under which two
subsequent buses pass a given point will display a Poissonian
distribution. This is a consequence of the absence of
correlations between the motion of different buses. Such situation
is, however, not welcomed by the drivers since then the
probability density that two buses arrive to a bus stop within
short time interval is large. In this case the first bus collects
all waiting passengers  and the second one that arrives bit later
will find the stop practically empty. This simple reasoning makes
clear that  existence of certain correlations between buses, that
will change the Poisson process, will be of favor. Indeed, in
Mexico various strategies have been developed to create such
correlations. Here we discuss the situation in three cities:
Cuernavaca, Puebla and Mexico City.\\

In Puebla there is not an additive mechanism that helps to
increase the bus correlation. Hence the headway statistics of
the buses in this city should be close to Poissonian. In Cuernavaca in
turn, the information about the times, when the buses pass certain
points, are notified and then sold to the bus drivers (there is
a real market with this information). The driver can in such a way
change the velocity of the bus in dependence on the position of
the bus in front of him and a bus-bus correlation appears. \\

To describe the correlations we use a modified version of the
celebrated Nagel-Schreckenberg cellular automaton (see
\cite{Helbing} and \cite{Chowdhury}). Consider $N$ equal cells on
the line (for our purpose we define the length of  one cell as 30
meters) and $n$ particles (buses) moving along it. Thus,
$c=\frac{n}{N}$ is the bus density and $\bar{d}=c^{-1}$ is the
mean distance between two neighboring buses. Furthermore, define
the maximal velocity $v_{max}$ of the bus and its probability $p$
to slow down $p \in [0,1].$ The Nagel-Schreckenberg model
describes the dynamics of the system with the help of the
following update rules. In time $t=0$ the positions $x_i$ of the
particle $i=1\ldots n$ are integer numbers randomly chosen from
the set $1\ldots N$ satisfying the condition $x_{i-1}>x_i$ for
every $i=2\ldots n.$ Furthermore, in time $t=0$ the velocity $v_i$
of the $i$-th particle is set to zero for  all $i:$ $v_i(0)=0$.
The buses start to move with velocity $v$ according to the update
rules that has to be applied simultaneously  to all particles.

\begin{itemize}
\item \underline{Step one:} When velocity of the bus  is smaller than a maximal velocity $v_{max},$
it  increases its velocity by one

\begin{equation}
v_i(t+1)=\min \left \{v_i(t)+1;v_{max} \right\} \label{rule1}
\end{equation}

\item \underline{Step two:} Particles with positive velocities are
randomly slowed down

\begin{equation}
v_i(t+1)=v_i(t+1)-1 \label{rule2}
\end{equation}

 with probability $p$

\item \underline{Step three:} Particles update their positions
according to

\begin{equation}
x_i(t+1)=\min \left \{x_i(t)+v_i(t+1);x_{i-1}(t+1)-1 \right \} \label{rule3},
\end{equation}

i.e., the particles move according to the rule $
x_i(t+1)=x_i(t)+v_i(t+1)$ with the restriction that they cannot
occupy the same cell or overtake each other. In that case the
particle hops to the cell behind the occupied one.

\end{itemize}

To create  a modification of the model and to adapt it to the
situation in Mexico we change the update rules on a subset $M$ of possible bus positions
where the information is passed to the driver. (The density of those points will be denoted
by $a$; $a=M/N$.)

To change the model we add to the steps
(\ref{rule1}),(\ref{rule2}),(\ref{rule3})
 an additive step (\ref{rule4}) that takes into account the processing of the information :

At the points $j \in M$ the information about the time interval
$\Delta t$ to a preceding bus, which passed that point, is
available to the driver. Using it the driver can modify the bus
velocity - to speed up if  $\Delta t>\bar{t}$ or to slows down for
$\Delta t <\bar{t}$. (Recall that $\bar{t}$ is the mean time
interval between subsequent buses.) Hence we change the model by
adding

\begin{itemize}
\item{\underline{Step four:}}

$$ v_j(t)=v_j(t) +1 $$

if $ t_i(j)- t_{i-1}(j)< \bar{t}$
\begin{equation}
v_j(t)=v_j(t) -1 \label{rule4}
\end{equation}
if $ t_i(j)- t_{i-1}(j)> \bar{t}$ where $ t_i(j)$ denotes the time when a bus
$i$ passed through the point $j$: $x_i(t_i(j))=j$.

\end{itemize}

Using the modified  model we focus on the $headway$ $statistics,$
i.e. on the probability distribution of the time intervals $\Delta
t$  between two subsequent buses passing a given point and compare
it with the results obtained in the cities. For the simulation of
the transport in Puebla we choose $c=1/40,\,v_{max}=2,p=0.5,$ and
$a=0$ taking into account the fact that there are no checking
points. It is not surprising that the headway statistics is in
this case  close to the Poisson distribution (see Figure 1)

\begin{equation}
P(t)=e^{-t} \label{Poisson},
\end{equation} when $t$ is the spacing re-scaled to the mean distance equal
1. Besides the headway statistics, we compare also a $number$
$variance$ $\Sigma^2(t)$ that is defined as

\begin{equation}
\Sigma^2(t)=\langle \left( n(t)-t \right)^2 \rangle \label{NV-definition},
\end{equation}

where $n(t)$ is the number of bus arrivals to a given point during
the time period of the length $t.$ Note that $\langle n(t)
\rangle=t$ due to the fact that $\langle t \rangle = 1.$ It can
easily be checked that for a Poissonian process

\begin{equation}
\Sigma^2(t)=t \label{NV_Poisson}.
\end{equation}

\inspicture

The number variance obtained from the data and from the simulation fits quite
well with this prediction (see Figure 3). The data show, however, a small
deviations from  (\ref{NV_Poisson}). This is a manifestation of a weak
interaction between the buses which probably originates from the fact that the buses
interact through the number of passengers waiting on the stops.
Namely when the distance between the buses is
large, more passengers are waiting in the stop and the delay of bus in the
stopping-place is longer.\\

To simulate the situation in Cuernavaca and in Mexico City we use parameters
$c=1/40,\,v_{max}=2,p=0.5,\,a=1/36,$ which represents one bus per 1.2
kilometers and one check point per one kilometer approximately. The modified
cellular automaton leads in this case to significant changes in the time
interval distribution (see Figure 2).  The distribution obtained from
the automaton fits well the observed time interval distribution. Moreover both distributions
conform well with the distribution (see Ref. \cite{Seba})

\begin{equation}
P(t)=\frac{32}{\pi^2}t^2 e^{\frac{4}{\pi}t^2} \label{Wigner},
\end{equation}

that describes  the spacing distribution of a Gaussian unitary ensemble of
random matrices (GUE). (It is known that this function describes also  the
distance distribution of certain one-dimensional interacting gases (see Ref.
\cite{Scharf}).

Similar agreement is observed also for the number variance (see Figure 3) where
GUE leads to

\begin{equation}
\Sigma^2(t)=\frac{1}{\pi^2}(\ln 2\pi t + \gamma +1) \label{NV_GUE}
\end{equation}

with $\gamma\approx0.57721566$ (see \cite{Mehta}).

\inspicture

However, as evident on the Figure 3, the interaction between the
buses in Cuernavaca and in Mexico City is stronger than that
resulting from  the cellular model and leads to the stronger
correlations. Whereas in the automaton model the correlations
exist between the nearest neighbors only, in Cuernavaca and Mexico
City one can recognize that interaction exists also between the
first, second and third neighbor. \\

We conclude that the modified Nagel-Schreckenberg cellular
automaton successfully describes microscopic properties of the bus
transport in some Mexican cities.  The velocity of the buses is
influenced by the information about their mutual positions so that
the drivers can optimize their rank in competing on the passengers
to be transported. This finally increase the coordination of the
bus motion and changes the time headway statistics.\\

Acknowledgement: This work has been supported by the grant
A1048101 and GACR 202/02/0088 of the Czech grant agency. The data
in the Mexican cities were collected with the friendly help of
Dr.Markus Mueller from the University in Cuernavaca and Dr.
Antonio Mendes-Bermudes from the University of Puebla. Authors are
also very grateful to Patrik Kraus from the University of Hradec
Kr\'alov\'e for entering the collected data into the computer.

\inspicture

\end{document}